\documentclass[prb,twocolumn,floatfix,preprintnumbers,amsmath,amssymb,superscriptaddress]{revtex4}
\usepackage{graphicx}
\usepackage{dcolumn}
\usepackage{bm}
\usepackage{color}
\usepackage{color}
\usepackage[latin1]{inputenc}
\usepackage[urlcolor=blue]{hyperref}
\hypersetup{backref, colorlinks=true} 
\begin{document}



\title{ {Spectral  characteristics and morphology of nanostructured Pb-S-O thin films synthesized via two different methods}}

\author{H.S.H. Mohamed}
\affiliation{Faculty of science, Fayoum University, 63514-Fayoum- Egypt.}
\affiliation{National research university, Krasnokazarmennaya 14, Moscow 111250, Russian Federation.}
\author{M. Abdel-Hafiez}
\affiliation{Faculty of science, Fayoum University, 63514-Fayoum- Egypt.}
\affiliation{Center for High Pressure Science and Technology Advanced Research, Shanghai, 201203, China}
\author{B. N. Miroshnikov}
\affiliation{National research university, Krasnokazarmennaya 14, Moscow 111250, Russian Federation.}

\author{I. N. Miroshnikova}
\affiliation{National research university, Krasnokazarmennaya 14, Moscow 111250, Russian Federation.}


\date{\today}

\begin{abstract}
Using two different experimental techniques, namely, chemical vapor deposition (CVD) and physical vapor deposition (PVD), we deposited a Lead sulphide (PbS) thin films with a very small lifetime $(10^{-9})$.  We investigated the morphology of the obtained PbS films using various techniques i.e.AFM, SEM, EDAX, AES and HRTEM . In the case of CVD, we found that the surface consists of grains with dimensions in the plane (diameter to 300\,nm and height up to 200\,nm), while the same order of the grain size has been observed for PVD. On the other hand, SEM investigation reveals that the PbS particles with various morphologies of both films have uniform and the particle size distribution. Small amount of Sodium was obtained from EDXS studies, which is may originate from the substrate where the deposition process has been produced at temperature 550-600$ ^\circ$C and for CVD at minimum accelerating voltage 5\,kV silicon are presented in the spectrum, which means that the region for X-ray generation voltage data exceeds the thickness of the films (where the thickness of films about 0.4 micron). AES confirm that the surface layer of these films (PVD) containing carbon and oxygen and it has a thickness of ~ 0.1$\mu$m. At a depth of 1.3 microns in films these elements is again increased, which corresponds to the film thickness of 1.5 $\mu$m. Layers of PVD films are seen by HRTEM and the studies confirm that oxygen-layer located on top of the structure, while the layers of CVD films have not only the oxygen along the crystallite boundaries, but also accumulate in the depth of the boundary with the substrate. Our results of morphology indicate that changing in spectral characteristics of films deposited by (CVD and PVD) is related to the structure and crystalline size.
\end{abstract}






\maketitle
\section{Introduction}

Solar cells technologies are becoming increasingly important in electric power generation. This is due to the fact that they provide more secure power sources and pollution free electric supplies. Semiconductor thin films are always in focus due to their outstanding electronic and optical properties and possible applications in various devices such as light-emitting diodes~\cite{VLC,FDM,MSM,KSP}, single electron transistors and field effect transistors~\cite{BAR}.The fabrication of Nanocrystalline semiconducting metal chalcogenide have drawn considerable interest in recent years,  because of their unusual optical and electric properties and potential applications in nanodevices~\cite{FDM2,MSM2,MMF}.In recent years, numerous efforts have been made to control the fabrication of nanostructured materials with various morphologies, since the novel properties and potential applications of nanomaterials depend largely on their shapes and sizes~\cite{MMF2,ZSG,MSN} . The electronic and optical properties of semiconductor materials can be changed by changing their size and shapes~\cite{PDY}. As an important IV-VI group semiconductor. PbS is an important direct narrow band gap semiconductor material ($\approx$ 0.41\,eV) with large excitation Bohr radius of 18\,nm~\cite{SS,SBP,SVL,FDM3}. Furthermore, it has been widely used in many fields such as Pb$^{2+}$ ion-selective sensors~\cite{HH}, IR detector~\cite{PG}, photography~\cite{PKN,MSN2}, and solar cell absorption~\cite{WCL,TKC,SKR}. Additionally, due to the non-linear optical properties of the PbS, it presents various applications in optical devices i.e., optical switch~\cite{RSK}. PbS has been utilized as photo resistance, diode lasers, humidity and temperature sensors, decorative and solar control coatings~\cite{PKN2,POP}. For these applications, we believe that further studies in order to explore the morphology of the PbS films with different particle sizes are required. It has been well reported that the particle diameters should vary from a few micrometers for infrared detector applications to several nanometers for quantum dots~\cite{RSM}. However, still intensive research interest in lead chalcogenides and various articles have been reported to explore new parameters in that system~\cite{DEA,MSN3}. Theoretical consideration has been reported by Neustroeva and Osipova in order to further understand the structures of PbS~\cite{RJC,RJC2}. Further properties of PbS layers on a silicon substrate with a sub-layer of SiO$_{2}$ have been studied ~\cite{VPE}. Although various studies show much attention to the structure, complexity, ambiguity, and manufacture~\cite{HES}, but still lack information from the physics point of view in these systems.


In the respect, this work aims to  explain that particular point and more information will be presented. One of the main motivation is to study the photosensitive films made from "Sapphire" because of its distinctive feature which indicated that the films made by the two methods (CVD and PVD). We demonstrate the changing in spectral characteristics of films, explain the effect of oxygen on the electronic structure of the energy bands and the morphology of nanostructured films Pb-S-O.
\section{Experimental}

Photosensitive films from PbS were synthesized by two different methods. First of all, physical method in this method it has been  used  metal spraying of PbS coating in a vacuum onto heated glass substrates and  subsequent heating these films  to 550-600$^{\circ }$C at atmospheric air pressure (PVD). Second method is "chemical" methods  in this method it has been  used PbS layer that has been precipitated from solutions is not subjected to heating above 100-120$^\circ$C, but an additional "oxidizing agent," frequently hydrate hydrazine, is added to the solutions (CVD)~\cite{LNK}.

The morphology and crystallography of the Pb-S-O polycrystalline thin films were studied by atomic force microscopy,  scanning the surface  and energy dispersive microanalysis (EDXS)and high resolution transmission electronmicroscopy (HRTEM).
\section{Results and discussion}

\begin{figure}
\includegraphics[width=22pc,clip]{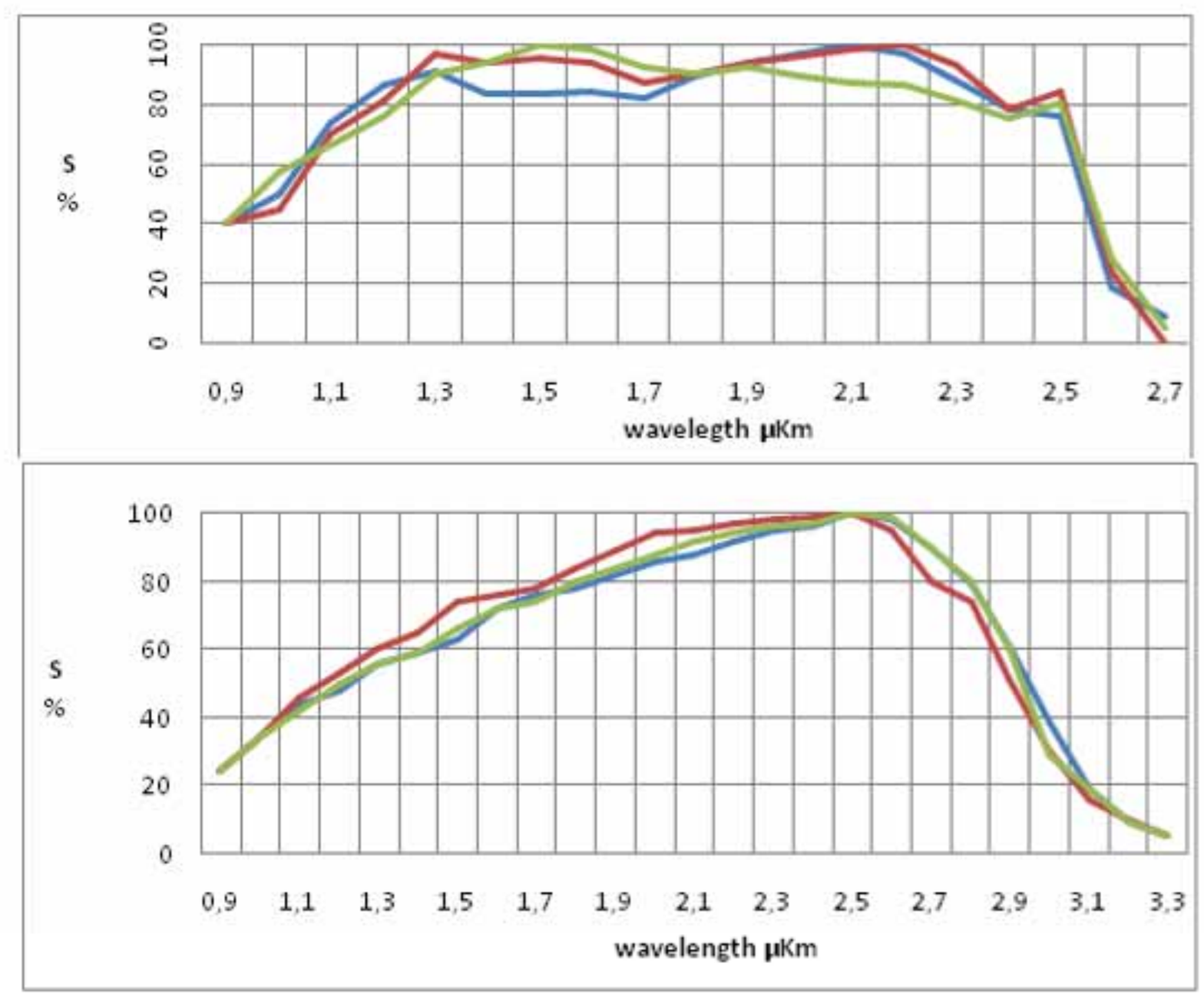}
\caption{\label{fig1} Spectral characteristics of photosensitive PbS films. (upper panel), films prepared by CVD and (lower panel), films prepared by PVD.}
\end{figure}

\begin{figure}
\includegraphics[width=18pc,clip]{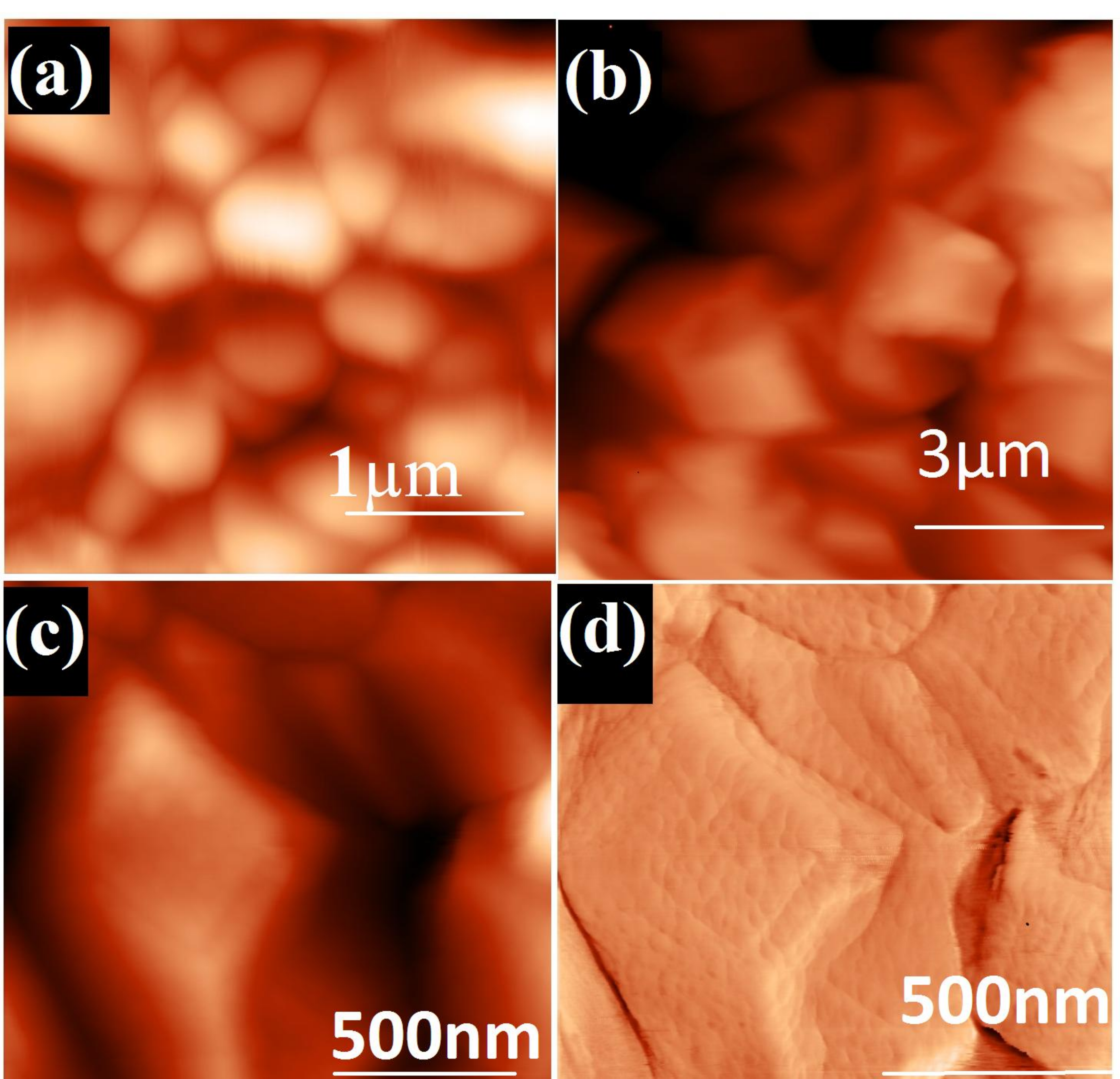}
\caption{\label{fig2} AFM picture of photosensitive PbS films (a) correspond to Scan the surface of the photosensitive layer produced by CVD, (b) correspond to Scan the surface of the photosensitive layer produced by PVD, (c) scan the PVD films in nanoscale  to obtain  the small grains in each cubic and (d) signal LF for PVD.}
\end{figure}
\subsection{Spectral characteristics of Pb-S-O}

The spectral characteristics of both PVD and CVD films have been investigated in~\cite{LNN,OAG,TNJ}. In the case of PVD (annealing process followed by heating) the the spectral characteristics have a peak near the wavelength, ($\lambda$) =2.5 microns and, corresponding to the bandgap ($E_{g}$ = 0.4eV at room temperature)as shown in (Fig.~\ref{fig1} lower panel). While for CVD films the spectral characteristics have several maximum, which are similar as the structures of the sandwich with different $E_{g}$ at different ($\lambda$) as shown in (Fig.~\ref{fig1} upper panel). From our morphological studies we can outline the difference between them, which may originate, first of all, from the non-optimal film thickness. The thickness of CVD films is about 0.4 $\mu$m, which is differ from the preferred thickness as 1.5$\mu$m but thickness for PVD is about 1.5$\mu$m. Secondly, the differences are due to the dissolving of oxygen inside the structure of PbS. It is necessary for dissolving oxygen in all sources, which will be joined with  sensitive layers, bonded with electrons and increase the life time of holes where the sensitization process by heat treatment in oxygen atmosphere was used in order to increase the photoconductivity of vacuum evaporated PbS thin films~\cite{EIA}. These holes are represented the majority carriers. However, it should be noted that the the effect of oxygen on the electronic structure of the energy bands, which brings in particular, a large decrease in the energy band gap under certain conditions.

Oxygen is an isoelectronic impurity, which replaces chalcogen in the lattice~\cite{WSW}. For such substitutional impurities, which introduce appreciable local distortions into the lattice, implies a strong interaction between the localized impurity states and extended states in the conduction band. As a result of presences  oxygen in PbS, even a small amount of an isoelectronic impurity gives rise to a splitting of the conduction band into two noncrossing subbands~\cite{NKM}. One of these subbands is formed of highly localized impurity states $E_{+}$, while the second subband is formed of extended conduction-band states $E_{-}$ that experience the effect of the narrow resonance band introduced by the isoelectronic impurity. In the literature, the localized state is positioned at 0.15\,eV from trapping levels and at 0.23\,eV from the recombination levels. So far, up to date, still lack of information which explain the reduction of energy gap by Urbach and the last years the role of oxygen in the structures of photosensitive materials , etc quite ambiguous where the oxygen defect plays an important role in the sensitive layers ~\cite{JWZ}. One can understand from the latter point that our spectral characteristics of the energy scale would be expected in  case of (CVD) and we expect the possibility of appearance effect Moss Burstein, which has been known in the structures based on InSb.

\subsection{Morphological and structure composition   obtained for Pb-S-O films}

PbS have attracted much attention in recent years due to their interesting morphology and potential applications~\cite{FDM2}. Spherical shapes of the CVD films revealed by AFM (Fig.~\ref{fig2}a). It consists of small grains with dimensions in the plane of the films (diameter to 300\,nm and height up to 200\,nm). The surface layers of the PVD films have a nearly parallelogram (pyramid) shapes as in (Fig.~\ref{fig2}b) each cube (parallelogram) consists of grains. These grains with dimensions are similar to those obtained for the CVD layer with the maximum sizes of grains are (500\,nm) as in (Fig.~\ref{fig2}c). In order to observe these grains, we must also watch the signal LF (Lateral Force, change the torsion of the cantilever deflection), which emphasize the features of relief as in (see Fig.~\ref{fig2}d). The average diameter of this grain is ranging from 40-60\,nm, while height of grain 3-5\,nm. The morphology of physical layers has distinct crystallographic faces and correct form, may be due to the high-temperature annealing 550-600$^{\circ }C $ in an uncontrolled atmosphere. The main disadvantage of this method is that the measurements are a shallow depth of field compared with the SEM. Based on these results, it is hardly difficult to say whether the grains in the relief are pyramid or parallelepiped. Therefore, our results cannot talk about the real shape of grains, but it have magnitude in order of 10 nanometers in the surface.
\begin{figure}
\includegraphics[width=18pc,clip]{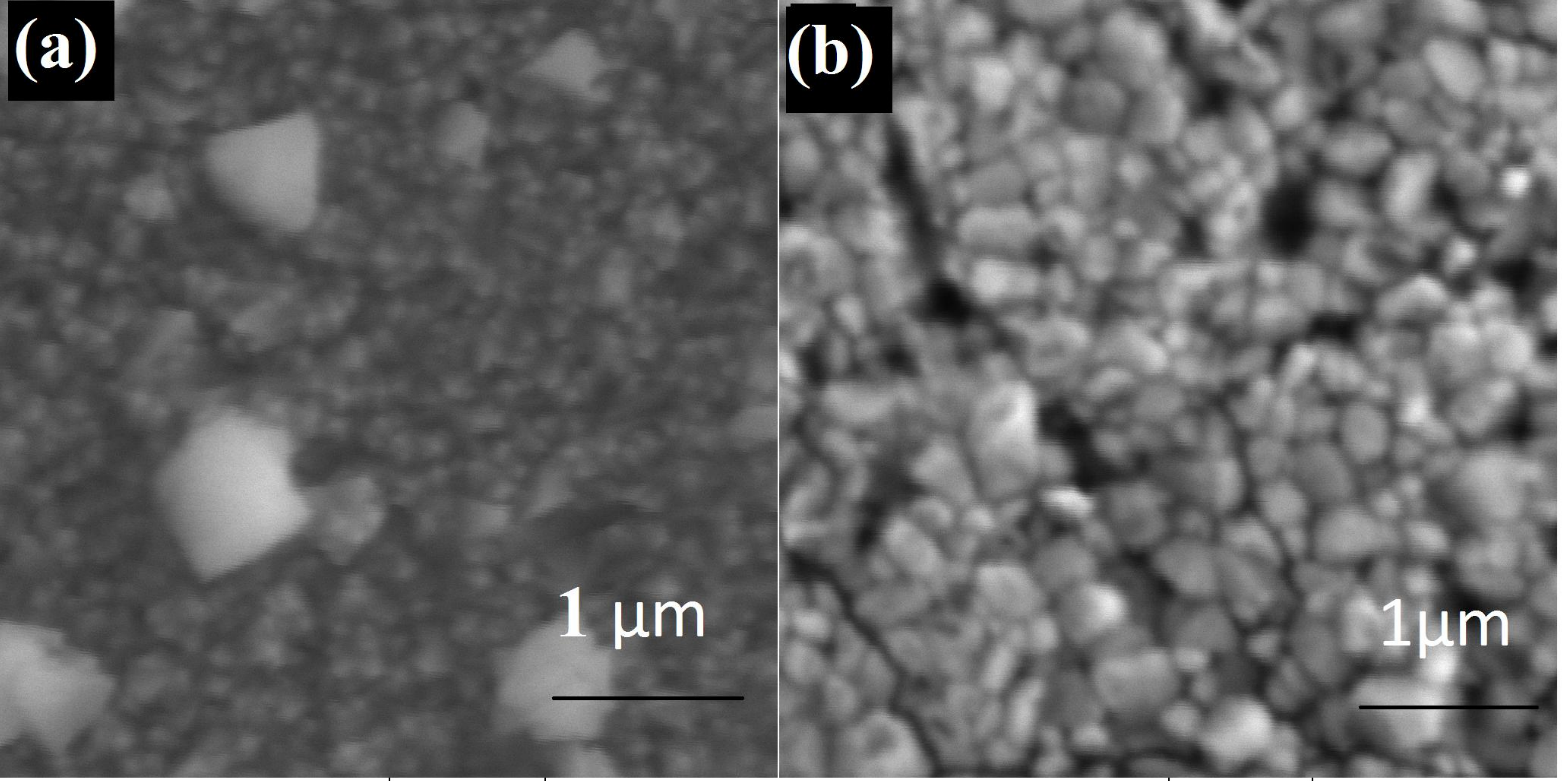}
\caption{\label{fig3} SEM picture of photosensitive PbS films. (a), films prepared by CVD and (b),films prepared by PVD.}
\end{figure}
The SEM investigation reveals that the PbS particles with various morphologies of both films have uniform and the particle size distribution. For PVD, the films  have polycrystalline form with grain sizes average  of 500\,nm and 600\,nm, which are in line with our results obtained from the AFM as shown in (Fig.~\ref{fig3}b). For CVD films, this method does not allow us to evaluate the smoothness of the surface, which must normally influence the spectral characteristics of photoresistsor as shown in figure.(see Fig.~\ref{fig3}a).
{\small
\begin{table*}
\caption{\label{tab:table 1}  {parameters and EDAX analysis for  all elements of both investigated CVD and PVD films.}}
\begin{tabular}{cccccccccc}
Method &Thickness ($\mu m$) &$R_{T}$($M\Omega$) & $E_{g}$(eV) & $V$(kV)$^{a}$\footnotetext[1]{We have used various accelerating voltage ranging between 5-30\,kV, but in this table we show only the 30\,kV for comparison.} & C($wt$\%) &  O($wt$\%) &Na/Si($wt$\%) &S($wt$\%) &Pb($wt$\%) \\

\hline
CVD &0.5 &0.34 &0.41 & 30 &69.83&17.83 &10.43& 0.97&0.85   \\
\hline
PVD &1.5  & 2.27&0.37& 30 &76.34&8.56 &1.12& 7.03& 6.85  \\
\end{tabular}
\end{table*}
}
\begin{figure}
\includegraphics[width=18pc,clip]{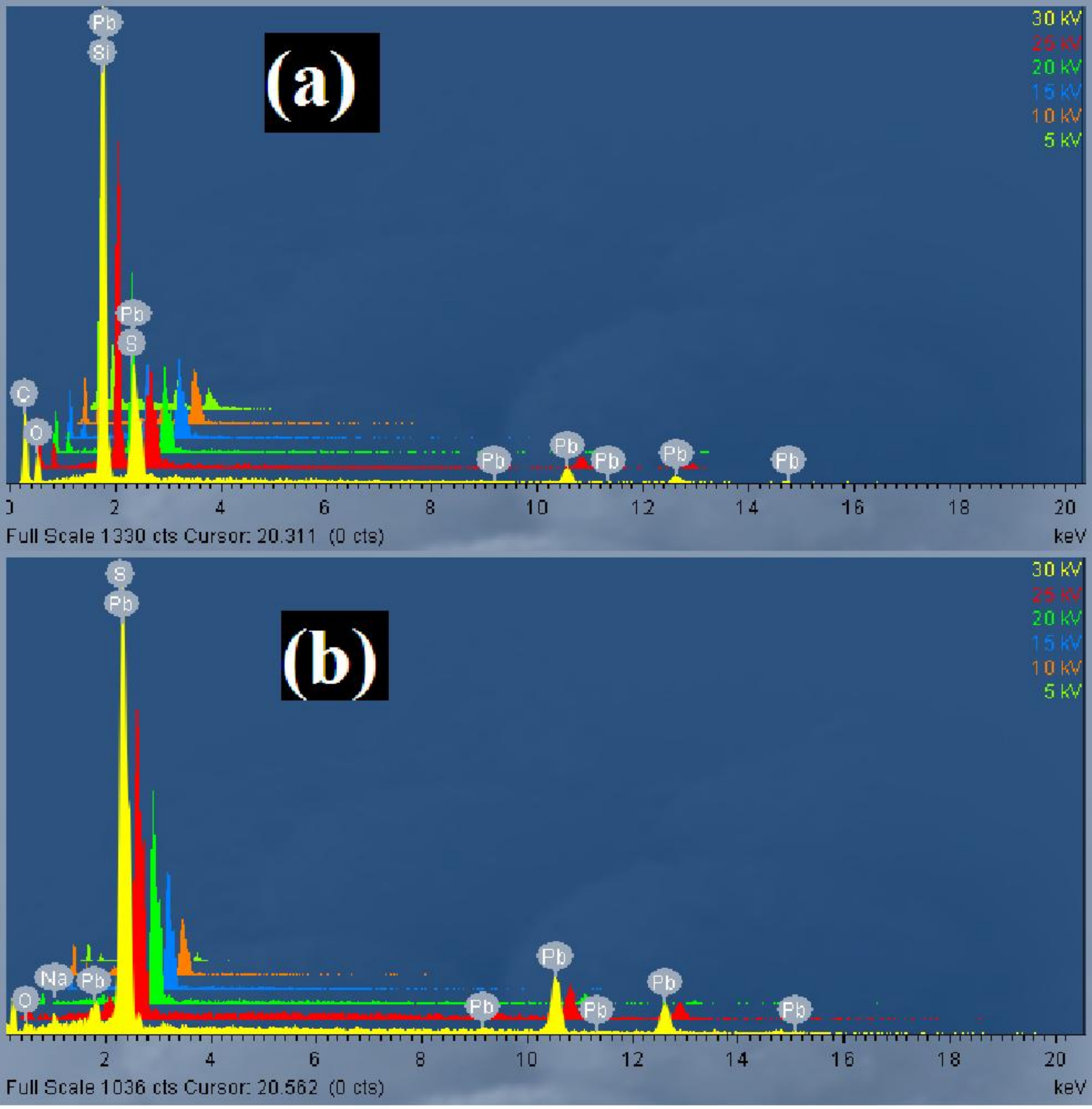}
\caption{\label{fig4} EDXS spectra of photosensitive PbS films. (a),films  prepared by CVD (a) and (b),films prepared by PVD.}
\end{figure}

\begin{figure*}
\includegraphics[width=40pc,clip]{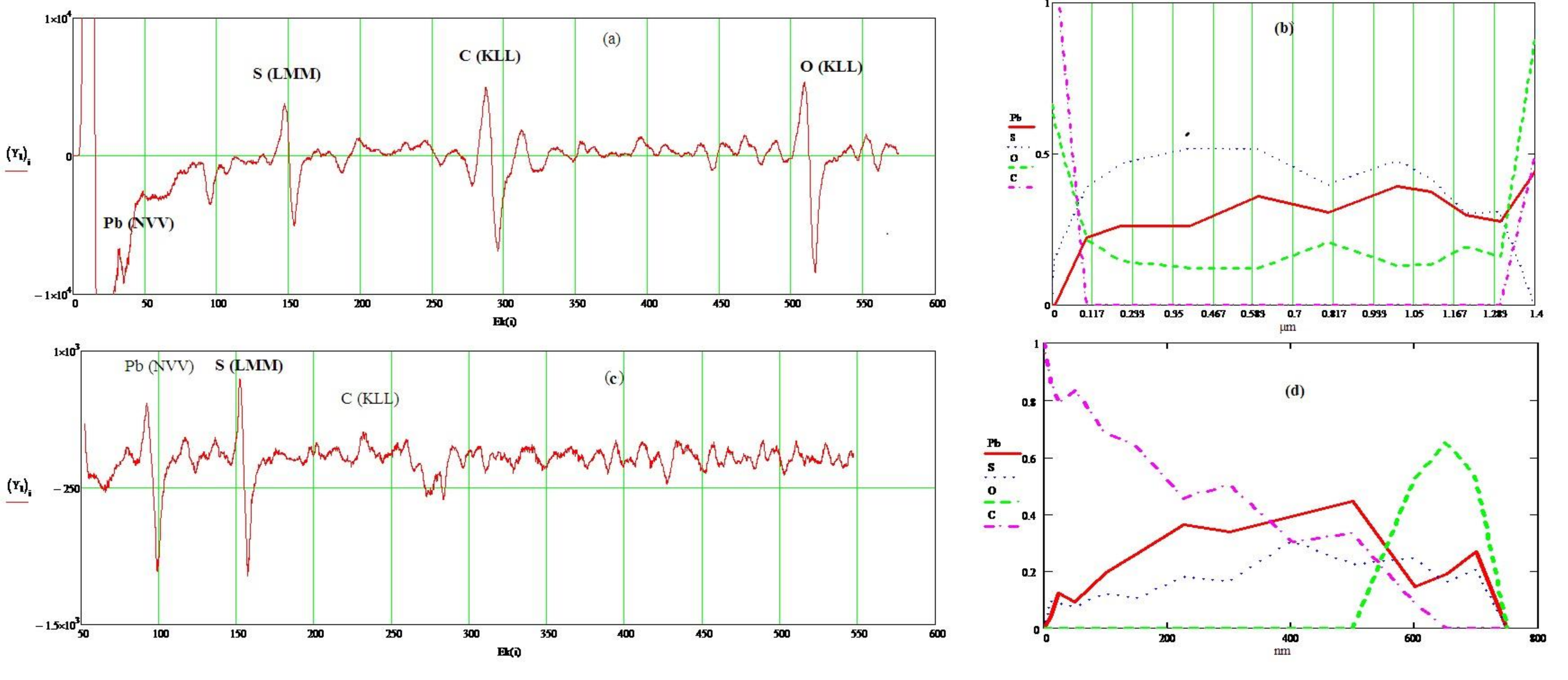}
\caption{\label{fig:wide} AES spectra of photosensitive PbS films. (a) correspond to Auger spectrum from samples produced by PVD,(b)show the ratio of elemental structure to depth of samples prepared by PVD,(c) correspond to Auger spectrum from samples produced by CVD,(d) show  the ratio of elemental structure to depth of samples  prepared by PVD.}
\end{figure*}

In the next step we detail the determination of chemical composition of the photosensitive layers by energy dispersive microanalysis (EDXS), which provides change in the depth of analyzed region of the layer. Typical spectra EDXS for samples prepared by "PVD" and "CVD" shown in ( (Fig.~\ref{fig4} and Table 1). EDXS studies reveals that films deposited by PVD existence of small amount of sodium   is about 0.4-1.6\,wt\% (see (Fig.~\ref{fig4}b) in the entire thickness of the film when accelerating voltage of electron beam  from 5\,kV to 30\,kV (which  the voltage of the beam determines the size of the "pear" in the  region of  X-ray emission).  Existing of sodium may be due to its diffusion from the substrate into the film, where the deposition process has been produced at temperature  550-600$^{\circ }$C~\cite{INM}. We accelerate voltage from 15\,kV to 30\,kV, taking into account necessary values voltage for more accurate to determine the concentration of lead, which $L_\alpha$ $\simeq$ 10.5\,keV. Also, throughout the film shows the presence of oxygen (20-25\,wt\%). Carbon present in the spectrum corresponds to the physical PbS hydrocarbon contamination unless the concentration of carbon not get involved in process technology layers, but the majority of carbon contamination became after the preparing process.

For films deposited by CVD in this process should be noted that even with a minimum accelerating voltage of 5\,kV present of silicon in the spectrum,(see (Fig.~\ref{fig4}a) which means that the region for X-ray generation voltage data exceeds the thickness of the films (0.4 micron). It is important to note that silicone, which is observed here comes from the substrate quartz ($SiO_2$),and  concentration ratio between Silicon and Oxygen is about 1:2 as we noted at($U> 20$)\,kV silicon concentration is constant$\simeq$10.5\,wt\% , and at ($U<10 $)kV $\simeq$ 5\,wt\%, where ($U> 20$ )kV X-ray radiation is generated mainly from the substrate, while ($U<10$)kV - from film. It is clearly that the existence of Oxygen in our films is not only  from the used substrate but also has been added to preparing processes. Thus, the average concentration of oxygen in samples prepared by (CVD) is three times smaller than in prepared by (PVD). Although it was unexpected, we had expected approximately equal value~\cite{INM2}. Therefore, the quantitative results are not conflicting with the data on the morphology, but it cannot provide dependable information on the distribution of oxygen in the photosensitive layers. For the sake of comparison, we have summarized the parameters of both investigated films extracted from EDAX analysis, see Table 1.


However, from the investigated results, it is difficult to determine in which form/areas that oxygen can be positioned,
we further used an Auger electron spectroscopy on LHS-10 (Leubold-Heraus, Germany) by using of coating argon layer. (Fig.~\ref{fig5}a) presents the differential Auger electron spectrum of "PVD" films. Auger peak shows that  Pb (NVV) has an energy of 96 eV, S(LMM) - 151 eV, C (KLL) - 275 eV and oxygen O (KLL) - 520 eV.  The  intensities of Auger peaks were standardized according to empirical sensitivity coefficients of corresponding Auger transitions. One clearly sees that the surface layer of these films (PVD) are containing carbon and oxygen with thickness of about ~ 0.1$\mu$m. We should also mention that around 1.3 microns depth in these films these elements are also increased, corresponds to the film thickness of 1.5 $\mu$m. In (Fig.~\ref{fig5}b) illustrates that at a depth of 0.8 and 1.2 microns, the oxygen ratio is significantly increasing. This fact indicting that the films are mainly consist of two grain chain of crystallites on the boundary which are inclusion oxygen.

On the other hand, the Oxygen ratio in CVD films is found to be less than in PVD films. Fig.~\ref{fig5}d shows the depth distribution of the detected elements and the thickness of these films have large carbon ratio. It is also necessary to note that this ratio is much smaller than in PVD films.
Depletion surface of the investigated films with sulfur are heated in the atmospheres, where the sulfur flies away from the surface, while at the substrate it is dissolved in the glass. The annealing process in high temperature leads to oxide a greater portion of the film, the oxide layer is formed at the edges of the crystallites which is accumulate between them. The distribution of oxygen in CVD films is localized at 200 nm from the substrate, which means that it is formed before depositing layer of PbS, which might be due to the compound $Pb(OH)_2$. Furthermore, the oxygen in CVD films substantially less than in PVD, see (Fig.~\ref{fig5}c).

Note that  the distribution of carbon in the thickness of the films is chemically deposited, with the growth of films by chemical deposition reaction mixture depleted in sulfur ions, increases the possibility of reactions with the formation of lead cyanide$ PbCN_2$ and lead acetate 2PbO.$Pb(CH_3COO)_2$.$H_2O$.

\begin{figure}
\includegraphics[width=18pc,clip]{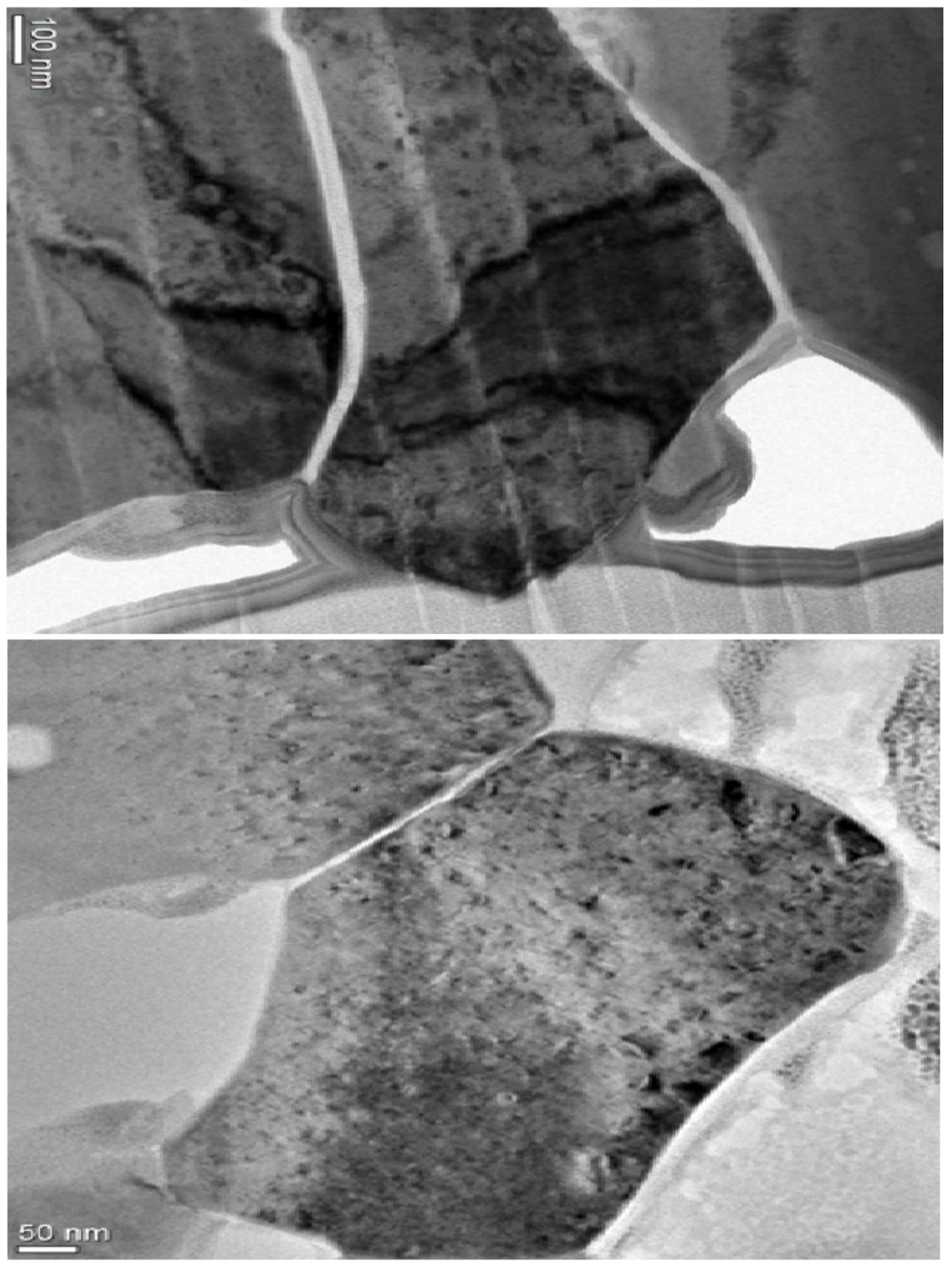}
\caption{\label{fig6}TEMHR picture of photosensitive PbS films. (upper panel), films prepared by CVD and (lower panel), films prepared by PVD.}
\end{figure}

In order to further explore the crystallographic investigations, we have used HRTEM. From these studies we noticed that both PVD and CVD films have Bright colors in the pictures belong to the light elements (primarily-oxygen) and a dark color reflects the atoms with higher numbers sulfur, lead). Clearly from (Fig.~\ref{fig6} lower panel we see the layering structure of the PVD films, where oxygen-layer is located on top of the structure and at the edges of large crystallites. On the other hand, for the CVD films upper panel (Fig.~\ref{fig6}, the oxygen is not only exists along the crystallite boundaries, but also accumulates in a depth on the boundary with the substrate. Thus, we can conclude that the thickness of films depends on the preparation technology and can determine a factor for the resistance value of the layers (which varies from tens of K$\Omega$ to several M$\Omega$. So, in order to improve any devices based on the PbS polycrystalline films, the difference in surface properties and sizes should be taken into account~\cite{INM3}.


\section{Summary}
In summary, we have used PVD and CVD to perform the PbS films. The investigations of morphological and spectra characteristic of PbS particles are presented. The elemental analysis of the obtained PbS particles were investigated by EDAX techniques. The fingerprint results of the PbS particles, prepared by both CVD and PVD, showed that both samples possess a cubic crystal structure, and considerable amount of oxygen in the  films, but they don`t have the same value. The ratio of oxygen on the metal-to-sulfur can change the spectral characteristics of PbS, the conductivity of the samples and energy gap, where oxygen can separate the conduction band into two sub-bands. Our results of morphology indicate that changing in spectral characteristics of films deposited by (CVD and PVD) is related to the structure and crystalline size.

This work is supported by the Egyptian-Russian governments. The work of MA was supported by HPSTAR grants and  The
authors would like to thank the anonymous referee for his
carefully reading the manuscript and for his valuable comments and
suggestions.

\end{document}